\documentclass[%
onecolumn
 reprint,
nofootinbib,
 amsmath,amssymb,
 aps, 
]{revtex4-2}
\usepackage[usenames, dvipsnames]{color}
\usepackage[autostyle]{csquotes}
\usepackage{hyperref}
\usepackage{color}
\usepackage{bigints}
\usepackage{graphicx}
\usepackage{dcolumn}
\usepackage{bm}
\usepackage{enumerate}
\usepackage{ulem}
\usepackage{float}
\usepackage{dcolumn}
\usepackage{bm}
\usepackage{placeins}
\usepackage{setspace}
\usepackage{amsmath}

\definecolor{MyDarkBlue}{rgb}{0,0.1,0.7}
\hypersetup{colorlinks,breaklinks=true,
  urlcolor={blue},citecolor={MyDarkBlue},linkcolor={Blue}}

\begin{document}


\title{Enhanced Gravitational Effects of Radiation and Cosmological Implications} 
\author{Hemza Azri$^{1,3}$}
\email{hm.azri@um.edu.my}
\author{Kemal G\"{u}ltekin$^{2}$}%
\email{kemalgultekin@iyte.edu.tr}
\author{Adrian K. E. Tee$^{1}$}%
\email{ad.astradrian@gmail.com}
\affiliation{$^{1}$Department of Physics, Faculty of Science, Universiti Malaya, Kuala Lumpur, 50603, Malaysia}
\affiliation{$^{2}$Department of Physics, \.{I}zmir Institute of Technology
G\"{u}lbah\c{c}e, Urla 35430, \.{I}zmir, T\"{u}rkiye}
\affiliation{$^{3}$National Centre for Particle Physics, Universiti Malaya, Kuala Lumpur, 50603, Malaysia}
\begin{abstract}
In the momentarily comoving frame of a cosmological fluid, the determinant of the energy-momentum tensor (EMT) is highly sensitive to its pressure. This component is significant during radiation-dominated epochs and becomes naturally negligible as the universe transitions to the matter-dominated era. Here, we investigate the cosmological consequences of gravity sourced by the determinant of the EMT. Unlike Azri and Nasri, Phys. Lett. B 836, 137626 (2023), we consider the most general scenario in which the second order variation of the perfect-fluid Lagrangian does not vanish. We analyze the dynamics of the power-law case and explore the cosmological implications of the scale-free model characterized by dimensionless couplings to photons and neutrinos. We show that, unlike various theories based on the EMT, the present setup — which leads to enhanced gravitational effects of radiation (EGER) — does not alter the time evolution of the energy density of particle species. Using current cosmological observations, we constrain the model parameters and show that EGER may offer a viable mechanism for alleviating the Hubble tension. Although it exhibits a phenomenological analogy to tightly-coupled relativistic fluid scenarios, EGER remains purely gravitational in origin and yields distinguishable signatures in the small-scale anisotropies of the cosmic microwave background. The radiation-gravity couplings we propose here are expected to yield testable cosmological and astrophysical signatures, probing whether gravity distinguishes between relativistic
and nonrelativistic species in the early universe.
\end{abstract}

\maketitle
\tableofcontents
\section{Introductory remarks and motivations}
\label{sec:intro}
General Relativity (GR) provides a consistent description of the gravitational phenomena and has passed numerous observational and experimental tests \cite{will, ishak}. However, several issues suggest that GR may not be the final theory of gravity. At the theoretical level, the presence of singularities (black hole and big-bang) and the lack of a consistent quantum formulation indicate fundamental limitations of the theory \cite{singular, qm}. From the observational side, the accelerated expansion of the universe, together with the recently emerged cosmological tensions within the standard model of cosmology, points to open questions that may require modifications of the gravitational interaction \cite{accel, hubbleT}.

A wide range of extensions to GR has been developed \cite{mog1, mog0}. Some of these modify the geometric part of the action by introducing curvature invariants beyond the Ricci scalar \cite{mog1, mog2, mog3}. Others modify the matter sector directly by incorporating explicit dependencies on the energy-momentum tensor (EMT) \cite{traceT,Tsquare,Tsquare2,Tsquare3}. Another line of investigation considers determinant-based actions. Determinants of rank-two tensors define scalar densities consistent with general covariance and have appeared historically in alternative formulations of gravity, such as the Eddington action \cite{affineinflation, edd}. Determinant structures involving the Ricci tensor and combinations with the metric determinant have been studied as possible extensions \cite{ricci_determinant, padmanhaban}. Recently, one of us  with a collaborator proposed an extension of matter-gravity coupling in GR, in which the determinant of the EMT, specifically the scalar $\bm{D}=|\textbf{det} \, T|/|\textbf{det}\, g|$ plays a central role \cite{azridet}. It was shown that $\bm{D}$ is highly sensitive to the pressure of the perfect fluid that describes an astrophysical object. As a consequence, significant deviations from the predictions of GR appear in compact objects such as neutron stars, where pressure is an essential component in the relativistic regime.

On the one hand, it is important to note that within the field-theoretic approach to GR, the Lagrangian contains the EMT as a source term coupled to the spin-2 field. Integrating out this gravitational degree of freedom induces effective interactions among the matter sources (EMT coupling terms) at the level of the Lagrangian \cite{EFTGR1,EFTGR2}. On the other hand, the determinant of the EMT arises naturally when constructing invariant terms. Indeed, the determinant of the spacetime metric tensor $\textbf{det} \, g$, required for maintaining diffeomorphism invariance of the gravitational action, is equivalent to the determinant of the “rescaled’’ EMT corresponding to the vacuum energy (cosmological constant) \cite{affineinflation}. Determinant structures of this type are well established in high-energy physics: the Nambu–Goto action for strings \cite{string} and the Born–Infeld action of electrodynamics \cite{born-infeld} are determinant-based. Likewise, Eddington- or Born–Infeld–inspired extensions of gravity employ determinant densities constructed from geometric and matter tensors \cite{banados}. Another principal motivation for considering the determinant of the EMT is that several well-known EMT–based models arise naturally from the expansion of $D$ around the vacuum. In fact, as $T_{\mu\nu} \rightarrow \mathcal{E}g_{\mu\nu} + T_{\mu\nu}$ where $\mathcal{E}\sim \Lambda_{\text{UV}}^{4}$ is the vacuum energy density in terms of an Ultra-Violet cutoff $\Lambda_{\text{UV}}$ (large values), this leads to 
\begin{eqnarray}
\bm{D}
\simeq
\mathcal{E}^{4}
\Bigg\{
1+
\left(\frac{1}{\mathcal{E}}\right)T
+\left(
\frac{1}{2\mathcal{E}}
\right)^{2}T^{2}
-\left(\frac{1}{2\mathcal{E}^{2}}\right)T_{\mu\nu}T^{\mu\nu}
 + \mathcal{O}\left(\frac{T}{\mathcal{E}}\right)^{3}
\Bigg\}.
\end{eqnarray}
In this paper, we revisit this determinant-based coupling framework and investigate its implications in a cosmological context. We propose an early-universe dynamics that operates entirely within the framework of the known particle content, which interacts with gravity minimally as in  GR, but is supplemented by additional generally invariant interaction terms constructed from the determinant of their EMT. We show that the determinant structure, being strongly pressure-sensitive, enhances the gravitational effect of radiation while leaving pressureless components unaffected, in contrast to trace- or quadratic-EMT couplings that generically alter both relativistic and nonrelativistic matter across all epochs \cite{traceT,Tsquare,Tsquare2}.

After deriving the gravitational field equations for the most general case involving an arbitrary function of the EMT determinant, we tackle the power-law models in a Friedmann-Lemaître-Robertson-Walker background and examine the associated continuity equations that govern deviations from the standard time evolution of radiation. We then focus on a scale-independent realization, in which the new radiation-gravity couplings are described by dimensionless parameters associated with the photon and neutrino sectors. We also derive the linear perturbation equations in the Newtonian gauge and track the deviations from standard radiation-gravity couplings. For this scale-independent scenario, we show that the redshift evolution of the radiation energy density coincides with the standard form. This result is notable, as it demonstrates that the new couplings dilute away analogously to standard cosmology, while still leading to an enhancement of the expansion rate. We show that the enhancement of the expansion rate remains consistent with the bounds from big bang nucleosynthesis. The allowed parameter space is constrained at the level of order ten percent, thus preserving the successful predictions of early-universe physics while permitting measurable deviations from the standard model during the radiation-dominated era.

To investigate the observational viability of the scale-free model of the novel radiation-gravity couplings, we carry out a Markov Chain Monte Carlo (MCMC) analysis using the most recent measurements of the cosmic microwave background (CMB), baryon acoustic oscillations (BAO), and Type Ia supernovae (SNeIa). We find that the new couplings display a close analogy to a tightly coupled relativistic fluids: at the background level, its effects align with those produced by shifts in $N_{\text{eff}}$ or by scenarios involving self-interacting dark radiation. However, this correspondence does not extend to the perturbations where the model provides distinct signatures in the small-scale CMB temperature anisotropies. These features provide a clear means of distinguishing the scale-free scenario of the proposed radiation-gravity couplings from conventional modifications to the radiation content. Moreover, the inferred parameters lead to a modest reduction in the Hubble tension. The size of this improvement is comparable to what is obtained in scenarios that introduce additional radiation; however, in the present case, the effect arises purely from the altered gravitational sector rather than from changes in the particle content of the early universe.

The paper is organized as follows. In Sec.~\ref{sec:detT}, we introduce the theoretical framework based on the determinant of the EMT and discuss its incorporation into the gravitational action. We then derive the corresponding cosmological background equations, including the expansion rate and the evolution of the energy densities, for the power-law class of models. In Sec.~\ref{sec:scale-free}, we focus on the scale-independent scenario, where we derive the linear perturbation equations and obtain analytic estimates of the parameter space relevant for addressing the Hubble tension. In Sec.~\ref{sec:analysis}, we present the results of the MCMC analysis and discuss them. Finally, Sec.~\ref{sec:conclusion} summarizes our findings and outlines future directions.

\section{Enhanced gravitational effect of radiation}
\label{sec:detT}
\subsection{The determinant of the stress-energy tensor and the gravitational action}
In this section, we introduce our gravitational framework which is based on the usual Einstein–Hilbert action of general relativity, minimally coupled to matter fields, and extended by the determinant of the EMT $T_{\mu\nu}$. The latter is defined as
\begin{eqnarray}
\label{det_t}
\textbf{det} \ T=\frac{1}{4!}
\epsilon^{\alpha\beta\gamma\rho}\epsilon^{\bar{\alpha}\bar{\beta}\bar{\gamma}\bar{\rho}}
T_{\alpha\bar{\alpha}}T_{\beta\bar{\beta}}T_{\gamma\bar{\gamma}}T_{\rho\bar{\rho}},
\end{eqnarray}
where $\epsilon^{\alpha\beta\gamma\rho}$ is the anti-symmetric Levi-Civita symbol. This determinant transforms identically to $\textbf{det}\, g$, and a physically meaningful quantity is then constructed from the ratio 
\begin{eqnarray}
 \bm{D}= \frac{|\textbf{det} \  T|}{|\textbf{det}\ g|}.   
\end{eqnarray}
The quantity $\bm{D}$ transforms clearly as a scalar function under general coordinate transformations. The generally invariant action involving the most general couplings from the determinant of the EMT is written as \cite{azridet}
\begin{align}
\label{general_action}
S =\int d^{4}x \sqrt{|\textbf{det} \  g|}
\,
\left\{
\frac{\left(R-2\Lambda\right)}{16\pi G} +\mathcal{L}[g]\right\}+\int d^{4}x \sqrt{|\textbf{det} \ g|}\,f(\bm{D}),
\end{align}
where $f(\bm{D})$ is an arbitrary function of $\bm{D}$.
An analogous formulation could also be implemented in the Palatini approach, where the geometric part of the action is written in terms of both the metric and an independent symmetric connection. In this paper, however, we will consider the standard metric formulation. The field equations are then obtained by performing a variation of the total action with respect to the metric tensor. The variation of the quantity $\bm{D}$ takes the form
\begin{eqnarray}
\label{deltaD}
    \delta \bm{D}= \frac{\delta |\textbf{det} \  T|}{|\textbf{det} \  g|} 
    + \bm{D}g_{\mu\nu}\delta g^{\mu\nu},
\end{eqnarray}
where the variation of the determinant of the EMT is given by
\begin{eqnarray}
    \delta |\textbf{det} \  T| = |\textbf{det} \  T| \left(T^{\text{inv}} \right)^{\mu\nu}
    \delta T_{\mu\nu},
\end{eqnarray}
where $(T^{\text{inv}})^{\mu\nu}$ is the inverse of the EMT.
Now we need to evaluate the right-hand side of this expression. Using the definition of the EMT in terms of the Lagrangian, $T_{\mu\nu}= \mathcal{L}g_{\mu\nu} -2\delta \mathcal{L}/\delta g^{\mu\nu}$, we get
\begin{eqnarray}
    \delta T_{\mu\nu}= \mathcal{L} \delta g_{\mu\nu}
    +\left\{\frac{1}{2}g_{\alpha\beta}\left(\mathcal{L}g_{\mu\nu}-T_{\mu\nu} \right)
    -2\frac{\delta^{2}\mathcal{L}}{\delta g^{\alpha\beta}g^{\mu\nu}} \right\}\delta g^{\alpha\beta}.
\end{eqnarray}
Finally
\begin{eqnarray}
    \left(T^{\text{inv}} \right)^{\mu\nu}\delta T_{\mu\nu}=
    -\left\{ \mathcal{L}\left(T^{\text{inv}}_{\mu\nu}-\frac{1}{2}g_{\mu\nu}T^{\text{inv}}\right)+ \frac{1}{2}T^{\text{inv}} T_{\mu\nu} \right\} \delta g^{\mu\nu} 
    -2\left(T^{\text{inv}} \right)^{\alpha\beta}\frac{\delta^{2}\mathcal{L}}{\delta g^{\alpha\beta} \delta g^{\mu\nu}}\delta g^{\mu\nu},
\end{eqnarray}
where $T^{\text{inv}}$ being the trace of the inverse of the EMT, and $T^{\text{inv}}_{\mu\nu}=g_{\alpha\mu}g_{\beta\nu}\left(T^{\text{inv}}\right)^{\alpha\beta}$.

All put together, the variation of the quantity $\bm{D}$ which is given by (\ref{deltaD}) takes the form
\begin{eqnarray}
    \delta \bm{D}= \bm{D} \left\{ 
    g_{\mu\nu} -\mathcal{L}\left(T^{\text{inv}}_{\mu\nu}-\frac{1}{2}g_{\mu\nu}T^{\text{inv}}\right)
    - \frac{1}{2}T^{\text{inv}} T_{\mu\nu}
    -2\left(T^{\text{inv}} \right)^{\alpha\beta}\frac{\delta^{2}\mathcal{L}}{\delta g^{\alpha\beta} \delta g^{\mu\nu}}
    \right\}\delta g^{\mu\nu}.
\end{eqnarray}
Using the above variations, the principle of least action applied to (\ref{general_action}) implies the gravitational field equations
\begin{eqnarray}
\label{gravitational_equations}
G_{\mu\nu}=
 -\Lambda g_{\mu\nu}
+\kappa T_{\mu\nu} 
+\kappa f(\bm{D})g_{\mu\nu}
+2\kappa \bm{D}f^{\prime}(\bm{D})\mathcal{T}_{\mu\nu}, 
\end{eqnarray}
where $G_{\mu\nu}$ is the standard Einstein tensor, $\kappa = 8\pi G$ (with $G$ being Newton's constant), and $f^{\prime}(\bm{D}) = df/d\bm{D}$. The tensor $\mathcal{T}_{\mu\nu}$ takes the form
\begin{eqnarray}\label{tau}
\mathcal{T}_{\mu\nu}= \ -g_{\mu\nu}
+\mathcal{L}\left(T^{\text{inv}}_{\mu\nu} -\frac{1}{2}g_{\mu\nu}T^{\text{inv}} \right)
+\frac{1}{2}T^{\text{inv}} T_{\mu\nu} + \ 2 (T^{\text{inv}})^{\alpha\beta}\frac{\delta^{2}\mathcal{L}}{\delta g^{\alpha\beta}\delta g^{\mu\nu}}.
\end{eqnarray}
It is worth noting that the quantity $D$ contains no derivatives of the metric and depends on it only algebraically through the EMT of the sources. Consequently, the gravitational field equations derived from the action remain second order in the metric, and the theory produces only the physical spin-2 degrees of freedom of GR. Since no higher-derivative curvature operators are introduced through the function $D$, the model avoids the appearance of Ostrogradsky instabilities or ghostlike modes.

Some care is required, however, when the EMT is sourced not by perfect fluids but by fundamental fields such as a scalar. Even in this case, the canonical EMT of a scalar field contains only first derivatives of the field, and the determinant $\bm{D}$ therefore introduces no second derivatives of either the metric or the scalar field. Nevertheless, in the present work we restrict attention to perfect fluids, which provide an excellent approximation for cosmological applications; within this setting the determinant structure is manifestly free from ghosts and other dynamical instabilities.

Before choosing the specific form for $f(D)$ to be studied here, it is worth examining the effect of the quantity $\bm{D}$ first. For a perfect fluid (a good approximation for a cosmological fluid) where $T_{\mu\nu}=(\rho + p)u_{\mu}u_{\nu}+pg_{\mu\nu}$ for each species, the determinant of the EMT, $\textbf{det} \ T \equiv \textbf{det} \ [T_{\mu\nu}]$, takes the form $\textbf{det} \ [g_{\mu\lambda}\,T_{\,\,\nu}^{\lambda}] = \textbf{det} \ g \times \textbf{det} \ \hat{T}$ where $\hat{T}$ is nothing but the matrix with the elements 
\begin{eqnarray}
\label{matrixT}
 \hat{T}_{\,\,\nu}^{\mu}=\left(\rho + p\right)u^{\mu}u_{\nu}
+p\delta_{\,\,\nu}^{\mu}.  
\end{eqnarray}
Hence, one gets $\bm{D} = |\textbf{det} \ \hat{T}|$. Now, in the momentarily inertial frame of the fluid, the calculation of the determinant of the matrix $\hat{T}_{\,\,\nu}^{\mu}$ is straightforward, and one finally gets
\begin{eqnarray}
\label{D_in_terms_of_rho_and_p}
\bm{D}= |\rho p^{3}|.
\end{eqnarray}

Therefore, in the comoving frame of the perfect fluid, $\bm{D}$ vanishes for baryons and cold dark matter (negligible pressure), ensuring that these species decouple from the new gravitational interaction we introduced. As a result, the coupling proportional to $\bm{D}$ active exclusively in radiation-dominated epochs, precisely when relativistic content governs the expansion history. Additionally, the whole structure is well-defined only when $\bm{D} \neq 0$, a condition that is required by the appearance of $T^{\mathrm{inv}}_{\mu\nu}$ in the field equations. Given its characteristics, we refer to this scenario as \textit{enhanced gravitational effects of radiation} (EGER).

\subsection{Power-law models and cosmological dynamics}
In analogy with extended gravity theories, the power-law structure is interesting on its own. One can consider models of the form $\bm{D}^{n}$ where the exponent $n$ is not necessarily an integer. Because the determinant itself carries a large mass dimension, making the action dimensionless requires introducing a constant with correspondingly high dimensionality. By introducing some constants $M_{\text{i}}$ with the dimension of mass, the general form of power-law models can therefore be written as
\begin{eqnarray}\label{D_i}
    f(\bm{D})=\sum_{\text{i}}M_\text{i}^{4(1-4n)}\bm{D}_{\text{i}}^n.
\end{eqnarray}
where we considered the contributions from various species $\text{i}$.

The gravitational equations (\ref{gravitational_equations}) involve the inverse of the EMT, $(T^{\text{inv}})^{\alpha\beta}$. For a perfect fluid, this is evaluated as follows. First, we write $(T^{\text{inv}})^{\mu\alpha}=g^{\alpha\nu}(T^{\text{inv}})^{\mu}_{\,\,\nu}$ and then determine the inverse of the matrix (\ref{matrixT}). Given a matrix of the form $A+UV^{\text{T}}$ where $A$ is a square invertible matrix and $U,V$ are column vectors, its inverse is given by the Sherman–Morrison formula \cite{matrix-inverse} 
\begin{eqnarray}
    \left(A+UV^{\text{T}} \right)^{-1}=
    A^{-1} - \frac{A^{-1}U\cdot V^{\text{T}} A^{-1}}{1+V^{\text{T}}A^{-1}U}.
\end{eqnarray}
For the case of a perfect fluid (\ref{matrixT}), $A=p_{\text{i}}I$ where $I$ is the $4\times 4$ identity matrix and $U=V=\sqrt{\rho_{\text{i}}+p_{\text{i}}}\, u$. By applying this to the above formula, one finally gets
\begin{eqnarray}
    (T^{\text{inv}})^{\mu\nu}=
    \frac{1}{p_{\text{i}}}
    \Biggl\{
    g^{\mu\nu}
    +\frac{(\rho_{\text{i}} + p_{\text{i}})}{\rho_{\text{i}}}u^{\mu}u^{\nu}
    \Biggl\}.
\end{eqnarray}
According to the previous discussion, this expression is not singular since it is valid only for $p_{\text{i}}\neq 0$ (relativistic species) whereas for $p_{\text{i}}=0$ (dust), the function $\bm{D}_{\text{i}}$ vanishes in the first place, and the structure tends to be the standard matter coupling of GR without any modification. By considering $\mathcal{L} = p_{\text{i}}$ for the Lagrangian of each fluid \cite{schutz_pf, fluid_lagrangian}, its second-order variation reads
\begin{eqnarray}
    \frac{\delta^{2}\mathcal{L}}{\delta g^{\mu\nu} \delta g^{\alpha\beta}}=\frac{1}{4}\left(\frac{1}{c_{\text{si}}^{2}} -1\right)\left(\rho_{\text{i}}+p_{\text{i}}\right)u_{\mu}u_{\nu}u_{\alpha}u_{\beta}
\end{eqnarray}
with $c_{\text{si}}^{2}=\delta p_{\text{i}}/\delta \rho_{\text{i}}$ (see \cite{nazari_second_derivative} for its derivation). Consequently, the presence of the second derivative of the Lagrangian through the equations of motion induces the adiabatic sound speed squared even at the background level. This was unjustifiably ignored when the determinant of the EMT was originally introduced \cite{azridet}. With these expressions at hand, we easily determine the tensor $\mathcal{T}_{\mu\nu}$ in (\ref{tau}) as 
\begin{eqnarray}
    \mathcal{T}_{\mu\nu} = 
    \frac{1}{2}
    \left(
    1+ \frac{p_{\text{i}}}{\rho_{\text{i}} }
    \right)
    \left(
    \frac{3\rho_{\text{i}}}{p_{\text{i}}} + \frac{1}{c_{s}^{2}}
    \right)u_{\mu}u_{\nu}.
\end{eqnarray}
All these put together, the gravitational field equations (\ref{gravitational_equations}) adapted to the power-law models take the form
\begin{eqnarray}
\label{eom_rho_p0}
G_{\mu\nu}= -\Lambda g_{\mu\nu}
+p_{\text{i}}\left(1 + M_\text{i}^{4(1-4n)}\frac{\rho_{\text{i}}^n}
{p_{\text{i}}^{1-3n}} \right)g_{\mu\nu}
+
(\rho_{\text{i}} + p_{\text{i}})
\left\{1 +n M_\text{i}^{4(1-4n)}\frac{p_{\text{i}}^{3n}}
{\rho_{\text{i}}^{1-n}}
\left(\frac{3\rho_{\text{i}}}{2p_{\text{i}}}+\frac{1}{2 c_{si}^2}
\right)
\right\}u_{\mu}u_{\nu} ,
\end{eqnarray}
where we took $\kappa=1$.

\label{sec:background}
\subsubsection{Friedmann and continuity equations of the power-law cases}
In what follows, the universe in its homogeneous approximation will be described by the Friedmann-Lem\^{e}tre-Robertson-Walker (FLRW) flat spacetime metric given by its line element 
\begin{eqnarray}
\label{flrw}
ds^{2}=-dt^{2}+a^{2}(t)d\vec{\textbf{x}}\cdot d\vec{\textbf{x}},
\end{eqnarray}
where $a(t)$ is the scale factor. Next, we will be interested in the energy evolution of the constituents of the universe which can be described by their energy density and pressure as the only relevant properties in the smooth background.

Applying the covariant divergence on the left-hand side of (\ref{eom_rho_p0}), and taking its time component ($\nu=0$), we derive the modified continuity equation
\begin{eqnarray}
\label{continuity0}
&&\dot{\rho}_{\text{i}}+3H(\rho_{\text{i}}+p_{\text{i}}) 
\nonumber\\
&&
+n M_\text{i}^{4(1-4n)}\rho_\text{i}^{4n-1}\left(\frac{p_{\text{i}}}{\rho_{\text{i}}}\right)^{3n}\left\{\left[4n\left(3+\frac{1}{c_{
si}^2}+
\frac{p_{\text{i}}}{\rho_{\text{i}}c_{si}^2}+\frac{3\rho_{\text{i}}}{p_{\text{i}}}
\right)-4\right]\dot{\rho}_{\text{i}}
+3H\left(
3+\frac{1}{c_{si}^2}+\frac{p_{\text{i}}}{\rho_{\text{i}}c_{si}^2}+\frac{3\rho_{\text{i}}}{p_{\text{i}}}
\right)\rho_{\text{i}}\right\}
=0, 
\end{eqnarray}
where $H=\dot{a}/a$ is the Hubble parameter, and  $p_{\text{i}}/
\rho_{\text{i}}=\omega_{\text{i}}$ is the constant equation of state
of the $\text{i}^{\text{th}}$ fluid component. Unlike the standard
continuity equation, we notice here the presence of the inverse of $
\omega_{\text{i}}$ which results from the inverse of the energy
momentum-tensor of radiation as we have mentioned previously. We notice
again that there are no effects from nonrelativistic matter where $
\omega_{\text{i}}=0$.

Assuming that the various species interact only gravitationally, the
continuity equation (\ref{continuity0}) holds for each type of
particles separately, namely, cold dark matter ($\text{i}=
\text{dm}$), baryons ($\text{i}=\text{b}$), photons ($\text{i}=
\gamma$) and neutrinos ($\text{i}=\nu$). Now, we adapt the gravitational field equation
(\ref{eom_rho_p0}) for the background metric (\ref{flrw}) and get the
expansion rate
\begin{eqnarray}
\label{fried1}
3H^{2}=\Lambda +
\sum_{\text{m}=\text{b,dm}}\rho_{\text{m}}
+
\sum_{\text{r}=\gamma,\nu}
\left[
\rho_{\text{r}}+M_{\text{r}}^{4(1-4n)}\left(\frac{1}{3}\right)^{3n}\left(16n-1\right)
\rho_{\text{r}}^{4n}
\right],
\end{eqnarray}
where we have used $\omega_{\text{b}}=\omega_{\text{cdm}}=0$ for baryons and cold dark matter species, $\omega_{\text{r}}=c_s^2=1/3$ for radiation, and have taken $u_{\mu}=(1, 0, 0, 0)$
for a comoving observer. Here, it is worth to note that the constants $M_{\text{r}}$ of mass dimension should not be confused with the masses of the relativistic species. 

The space-space components of the field
equations (\ref{eom_rho_p0}) lead to the time change of the Hubble parameter as
\begin{eqnarray}
\label{fried2}
\dot{H}=-\frac{1}{2}\sum_{\text{m}=\text{b,dm}}\rho_{\text{m}}
-
\sum_{\text{r}=\gamma,\nu}\left[
\frac{2}{3}\rho_{\text{r}}+8 n M_{\text{r}}^{4(1-4n)}\left(\frac{1}
{3} \right)^{3n}
\rho_{\text{r}}^{4n}\right].
\end{eqnarray}
Returning to the continuity equation (\ref{continuity0}), since the quantity $\bm{D}$ vanishes for nonrelativistic matter, the time evolution of the latter is not affected by the new
interaction terms, thus $\dot{\rho}_{\text{b},\text{cdm}}
+3H\rho_{\text{b},\text{cdm}}=0$, and in terms of the redshift $z$
one has $\rho_{\text{b},\text{cdm}}=\rho_{0\text{b},\text{cdm}}
(1+z)^{3}
$. For photons and (relativistic) neutrinos, it reads
\begin{eqnarray}
\label{continuityTheta}
\left(\frac{\rho_\text{r}^{1-4n}+\Theta_1}{\rho_\text{r}^{1-4n}+\Theta_2}\right)
\frac{d\text{ln} \rho_\text{r}}{dt}+4\frac{d\text{ln}a}{dt}=0,\label{rd}
\end{eqnarray}
where
\begin{align}
&\Theta_1=4 n M_\text{r}^{4(1-4n)}\left(\frac{1}
{3} \right)^{3n} \left(16n-1\right), \\   
&\Theta_2=12 n M_\text{r}^{4(1-4n)}\left(\frac{1}
{3} \right)^{3n}.
\end{align}
It is clear that the time evolution of relativistic species generally differs from that of standard cosmology $\rho_{\text{r}} \sim (1+z)^{4}$. However, it should be noted that the evolution becomes identical to the standard case, i.e. unaffected by the modification when $\Theta_{1} = \Theta_{2}$, a condition satisfied by the scale-independent model ($n = 1/4$), which we examine in the next section.
\section{Cosmological implications of the scale-free model}
\label{sec:scale-free}
\subsection{Background evolution and big bang nucleosynthesis constraints}
According to the expression (\ref{D_i}), the scale-independent construction arises for $n=1/4$ or
\begin{eqnarray}
 f(\bm{D})=\sum_{\text{i}} \lambda_{\text{i}}\,\bm{D}_{\text{i}}^{1/4},   
\end{eqnarray}
where $\lambda_{\text{i}}$ are dimensionless constants referring to the couplings of various species.
A reason for choosing a scale-free model is that it carries the same mass dimension as the fluid energy density itself. As a result, the theory requires no additional mass-scale, and the only energy scales appearing in the setup are those already encoded in the physical fluid variables (energy density and pressure). Therefore, the strength of the new coupling is controlled solely by dimensionless parameters associated with each relativistic species.

Again, the preceding analysis shows that a non-vanishing determinant implies that the modification affects only the radiation sector. This forces the non-relativistic matter to detach from these couplings. Therefore, the novel contribution targets only the radiation sector which will be described by the free parameters $\lambda_{\text{r}}=\lambda_{\gamma}, \lambda_{\nu}$ for photons and relativistic neutrinos respectively. For this model, the Friedmann equations (\ref{fried1})-(\ref{fried2}) take the form
\begin{eqnarray}
\label{fried11}
3H^{2}=
\rho_{\text{m}} 
+
\sum_{\text{r}=\gamma, \nu}
\left(
1+3^{1/4}\lambda_{\text{r}}
\right)\rho_{\text{r}} 
+ \Lambda ,
\end{eqnarray}
\begin{eqnarray}
\label{fried22}
\dot{H}=
-\frac{1}{2}\rho_{\text{m}}
-\frac{2}{3}
\sum_{\text{r}=\gamma, \nu}
\left(
1+3^{1/4}\lambda_{\text{r}}
\right)\rho_{\text{r}}.
\end{eqnarray}
Again, as in standard cosmology $\rho_{\text{m}}$ involves both baryons and cold dark matter energy densities whilst radiation, encoded in $\rho_{\text{r}}$, involves photons (and $e^{+}e^{-}$ pairs when prior to big bang nucleosynthesis) and possibly, three flavors of left-handed neutrinos as described by the SM of particle physics. 
Despite the complexity of the gravitational field equations (\ref{gravitational_equations})-(\ref{tau}), the cosmological equations (\ref{fried11})-(\ref{fried22}) reveal a simple but key consequence: the present setup leads to effective gravitational couplings that differ between matter and radiation. While pressureless matter (modeled as dust) continues to gravitate with the standard Newton constant $G$, radiation experiences a rescaled coupling of the form $(1+3^{1/4}\lambda_{\text{r}})G$. The values of the coupling parameters $\lambda_{\text{r}}$ assigned to each relativistic species determine their influence on key cosmological quantities, such as the Hubble parameter and the sound horizon. 

On the other hand, the continuity equations (\ref{continuityTheta}) reduce to their standard form for this model ($n=1/4$). Consequently, the solution is given by $\rho_{\text{r}}=\rho_{\text{r}_{0}}(1 + z)^{4}$ in terms of the redshift $z$. This feature is central to the mechanism by which the enhanced gravitational coupling effectively tracks the radiation component and naturally dilutes as the universe transitions to the matter-dominated phase. As we shall discuss later, an interesting implication of this behavior is that the increase in $H(z)$ prior to recombination reduces the sound horizon and raises the CMB-inferred value of $H_0$, which may contribute to easing the Hubble tension.

In a broad class of scenarios beyond the standard model of cosmology or particle physics (if new particle species are involved), departure from the the standard dynamics is conveniently described in terms of an effective expansion rate $H^{\prime}$, related to the standard Hubble rate $H$ through a dimensionless factor $S$ as $H\rightarrow H^{\prime}=SH$. It has been shown that analytic fits to big bang nucleosynthesis (BBN) imply that for non-standard expansion rate $SH$ which might arise generally from new physics must satisfy $0.85\le S \le 1.15$ \cite{bbn}. In the EGER, deviations from the standard case $S=1$ arise from the dimensionless couplings $\lambda_{\text{r}}$ as $S=\left(1+3^{1/4}\lambda_{\text{r}}\right)^{1/2}$ according to (\ref{fried11}), and therefore $-1.1\times 10^{-1} \leq \lambda_{\text{r}}\leq 1.1\times 10^{-1}$. These bounds show that the EGER remains tightly constrained by big bang nucleosynthesis, with the free parameters limited to values of order one tenth. The result ensures that the scenario preserves the successful predictions of early-universe physics while still allowing for measurable deviations from the standard model in the radiation-dominated era.

\subsection{Linear scalar perturbations}
\label{sec:perturbations}

In this section, we will derive the scalar perturbations of the scale-independent model of the EGER. We will work in conformal-Newtonian gauge and write our perturbed metric as
\begin{eqnarray}
ds^{2}=
a^{2}(\eta)\left[ -(1+2\Psi(\vec{\textbf{x}},t))d\eta^{2}
+(1-2\Phi(\vec{\textbf{x}},t))
d\vec{\textbf{x}} \cdot d\vec{\textbf{x}}
\right].
\end{eqnarray}
Here  $\eta$ is the conformal time, $\Psi$ is the gravitational potential from which the Newtonian gravity is recovered at scales smaller than the Hubble radius. The function $\Phi$ represents a local distribution of the scale factor. For perfect fluids, one immediately has $\Phi=\Psi$. Additionally, the speed of sound reads $c_{s}^{2}=\bar{p}/\bar{\rho}=\delta p/\delta \rho$ where $\bar{\rho}$, $\bar{p}$ are the background quantities whereas $\delta \rho$ and $\delta p$ are the perturbation quantities. In addition, one writes the fluid velocity perturbation as $u^{\mu}=a^{-1}\delta_{0}^{\mu}+\delta u^{\mu}$ in which $\delta u^{i}=v^{i}$ is a small velocity. From the latter, one defines the scalar degree of freedom (velocity divergence) $\theta=\vec{\nabla} \vec{v}$. On the other hand, since the particle species are approximated by perfect fluids, then no anisotropic stresses are considered. Therefore, the perturbations are totally described by only the two degrees of freedom, $\delta$ and $\theta$. 

From the gravitational field equations (\ref{eom_rho_p0}), and for the scale-independent model ($n=1/4$), one writes a total (an effective) EMT involving the EGER corrections as
\begin{eqnarray}
\label{totalemt}
T^\mu_{\text{tot}\,\, \nu}= p\left(1 + \lambda
\left(\frac{\rho}{p}\right)^{1/4} \right)
\delta^\mu_{\ \nu}
+
(\rho + p)
\left\{1 +\frac{\lambda}{2}\left(\frac{p}{\rho}\right)^{3/4}
\left(\frac{3\rho}{2p}+\frac{1}{2 c_{s}^2}
\right)
\right\}u^\mu u_\nu ,
\end{eqnarray}
where we neglected the cosmological constant term. Now, we consider linear perturbations for the energy density and pressure about the background as
\begin{eqnarray}
\rho=\bar{\rho}+ \delta \rho, \quad
p=\bar{p}+ \delta p
\end{eqnarray}
for various species, and define the dimensionless perturbation  $\delta=\delta \rho/\bar{\rho}$ which describes the relative deviation of the energy density from the mean background density. For the cosmological perturbation equations, we will use almost the same notation of Ref. \cite{ma_bertschinger} for the main variables. To linear order in the perturbations, the components of this EMT read 
\begin{eqnarray}
    &&T^{0}_{\text{tot}\,\,0}=
    -(\bar{\rho} + \delta \rho) + \tilde{T}^{0}_{\,\,0}, \\
    &&T^{0}_{\text{tot}\,\,i}=
    (\bar{\rho} + \bar{p})v_{i} + \tilde{T}^{0}_{\,\,i} ,\\
    &&T^{i}_{\text{tot}\,\,j} =
    (\bar{p} + \delta p)\delta^{i}_{\,\,j}
    + \tilde{T}^{i}_{\,\,j} + \tilde{\Sigma}^{i}_{\,\,j},
\end{eqnarray}
where the first contributions are the standard terms that arise in standard cosmology, and the last terms are given by 
\begin{align}
& \tilde{T}^0_{\ 0}=-\frac{\lambda}{4}\left(\frac{\bar{p}}
{\bar{\rho}}\right)^{3/4}\left(\frac{3\bar{\rho}}{\bar{p}}+
\frac{\bar{p}}{\bar{\rho}c^2_{\text{s}}}+1-\frac{1}{c^2_{\text{s}}}\right)\bar{\rho}\nonumber\\ & \ \ \ \ \ \ \ \  +\frac{\lambda}{16}\left(\frac{\bar{p}}
{\bar{\rho}}\right)^{3/4}\left(\frac{3\bar{p}}{\bar{\rho}c^2_{\text{s}}}-
\frac{15\bar{\rho}}{\bar{p}}+1-\frac{1}{c^2_{\text{s}}}\right)\delta\rho
+\frac{\lambda}{16}\left(\frac{\bar{p}}{\bar{\rho}}\right)^{3/4}\left(\frac{3\bar{\rho}^2}{\bar{p}^2}+\frac{3\bar{\rho}}{\bar{p}}-\frac{7}{c^2_{\text{s}}}-\frac{3\bar{\rho}}
{\bar{p}c^2_{\text{s}}}\right)\delta p , \\
& \tilde{T}^0_{\ i}=\frac{\lambda}{4}\left(\frac{\bar{p}}{\bar{\rho}}
\right)^{3/4}\left(\frac{3\bar{\rho}}{\bar{p}}+\frac{\bar{p}}
{\bar{\rho}c^2_{s}}+3+\frac{1}{c^2_{s}}\right)\bar{\rho}v_i , \\
&\tilde{T}^i_{\ j}=\lambda\left(\frac{\bar{p}}{\bar{\rho}}
\right)^{3/4}\bar{\rho}\delta^i_{\ j}+\frac{\lambda}
{4}\left(\frac{\bar{p}}{\bar{\rho}}\right)^{3/4}\left(\delta \rho +3\frac{\bar{\rho}}{\bar{p}}\delta p
\right)\delta^i_{\ j},
\end{align}
where $\lambda$ is the dimensionless constant characterizing the coupling to the EMT in action (\ref{general_action}). Needless to say, the terms involving $\lambda$ contribute to relativistic species (radiation) only. The tensor $\tilde{\Sigma}^{i}_{\,\,j}$ is the total anisotropic
stress of the fluid, that is, $\tilde{\Sigma}^{i}_{\,\,j}
=T^i_{\text{tot}\ j}-\delta^{i}_j T^k_{\text{tot}\ k}/3$. Here, the terms proportional to $\bar{\rho}/\bar{p}$, i.e. the inverse of the equation of state, are generated from varying the determinant of the EMT.

The evolution equation for the gravitational scalar potentials reads
\begin{align}
\label{deltaT00}
&k^2\Phi+3\mathcal{H}\left(\Phi^\prime +
\mathcal{H}\Psi\right)=4\pi G a^2 \delta T^0_{\ 0}+4\pi G a^2 \delta
\tilde{T}^0_{\ 0}, \\
&k^2\left(\Phi^\prime+\mathcal{H}\Psi\right)=4\pi G a^2
\left(\bar{\rho}+\bar{p}\right)\theta+\lambda\pi G
a^2\left(\frac{\bar{p}}{\bar{\rho}}
\right)^{3/4}\left(\frac{3\bar{\rho}}{\bar{p}}+\frac{\bar{p}}
{\bar{\rho}c^2_{\text{s}}}+3+\frac{1}{c^2_{\text{s}}}\right)\bar{\rho}\theta
\label{deltaTii}, \\
&\Phi^{\prime \prime}+\mathcal{H}\left(\Psi^\prime+2\Phi^\prime
\right)+\frac{k^2}{3}\left(\Phi-\Psi\right)+\left(2\mathcal{H^
\prime}+\mathcal{H}^2\right)\Psi=\frac{4\pi}{3}G a^2 \delta T^i_{\
i}+\frac{4\pi}{3}G a^2 \delta \tilde{T}^i_{\ i}, \\
&k^2\left(\Phi-\Psi\right)=12 \pi G a^2\left(\bar{\rho}+\bar{p}
\right)\tilde{\sigma},
\end{align}
where we have introduced $\delta T^0_{\ 0}=-\delta \rho$, $\delta T^i_{\ i}=3 \delta p$ and
\begin{align}
&\delta\tilde{T}^0_{\ 0}=\frac{\lambda}{16}\left(\frac{\bar{p}}
{\bar{\rho}}\right)^{3/4}\left(\frac{3\bar{p}}{\bar{\rho}c^2_{\text{s}}}-
\frac{15\bar{\rho}}{\bar{p}}+1-\frac{1}{c^2_{\text{s}}}\right)\delta\rho
+\frac{\lambda}{16}\left(\frac{\bar{p}}{\bar{\rho}}\right)^{3/4}\left(\frac{3\bar{\rho}^2}{\bar{p}^2}+\frac{3\bar{\rho}}{\bar{p}}-\frac{7}{c^2_{\text{s}}}-\frac{3\bar{\rho}}
{\bar{p}c^2_{\text{s}}}\right)\delta p,  \\
&\delta\tilde{T}^i_{\ i}=\frac{3\lambda}
{4}\left(\frac{\bar{p}}{\bar{\rho}}\right)^{3/4}\left(\delta \rho +3\frac{\bar{\rho}}{\bar{p}}\delta p
\right), \\
&\left(\bar{\rho}+\bar{p}\right)\tilde{\sigma}\equiv-\left(1+\lambda\left(\frac{\bar{\rho}}{\bar{p}}\right)^{1/4}\right)\left(\hat{k}_i
\hat{k}^j-\frac{1}{3}\delta^j_{\ i}\right)\Sigma^i_{\ j}
\end{align}
with $\Sigma^i_{\ j}$ being the anisotropic stress of the fluids. Applying the covariant conservation law (arising from the Bianchi identity) on the total EMT (\ref{totalemt}), and working in the Fourier $k$-space, we obtain the Euler and the continuity equations as
\begin{align}
&\delta^\prime+\frac{b}{a}3\mathcal{H}\left(\frac{\delta p}{\delta
\rho}-\omega\right)\delta+\frac{c}{a}(1+\omega)(\theta-3\Phi^
\prime)=0, \\
&\theta^{\prime}+\mathcal{H}\left(1-3\omega \frac{d}{f}\right)\theta-\frac{e}{c}
\frac{\delta p / \delta \rho}{(1+\omega)}k^2 \delta + \frac{1}{c}
k^2\tilde{\sigma}- k^2 \Psi=0,
\end{align}
with the following coefficients
\begin{align}
&a=1-\frac{\lambda}{16}\omega^{3/4}\left[\left(\frac{3\omega}
{c^2_{s}}-15\omega^{-1}+1-\frac{1}{c^2_{s}}\right)+
\left(3\omega^{-2}+3\omega^{-1}-\frac{7}{c^2_{s}}-
\frac{3\omega^{-1}}{c^2_{s}}\right)\frac{\delta p }{\delta \rho}
\right], \\
&b=\frac{1+\frac{\lambda}{16}\omega^{3/4}\left(27\omega^{-1}-
\frac{3\omega}{c^2_{s}}-1+\frac{1}{c^2_{s}}\right)+
\frac{\lambda^2}{4}\omega^{3/2}\left(3\omega^{-2}-\frac{1}{c^2_{s}}
\right)}{1+\frac{\lambda}{4}\omega^{3/4}\left(3\omega^{-1}+
\frac{\omega}{c^2_{s}}-1+\frac{1}{c^2_{s}}\right)}, \\
&c=1+\frac{\lambda}{4}\omega^{3/4}\left(3\omega^{-1}+\frac{\omega}
{c^2_{s}}+3+\frac{1}{c^2_{s}}\right)(1+\omega)^{-1},\\
&d=1+\lambda \omega^{-1/4},\\
&e=1+\frac{\lambda}{4} \omega^{3/4}\left(\left(\frac{\delta p}{\delta
\rho}\right)^{-1}+3 \omega^{-1}\right) ,\\
&f=1+\frac{\lambda}{4}\omega^{3/4}\left(3\omega^{-1}+\frac{\omega}
{c^2_{s}}-1+\frac{1}{c^2_{s}}\right).
\end{align}
Since we consider $\dot{\omega}=0$ for the equation of state, that is $\delta p/\delta
\rho=c^2_{s}=\omega$, we get $a=c=d=e=f=1+\lambda \omega^{-1/4}$ which simplifies the equations for $\delta$ and $\theta$ as
\begin{align}
&\delta^{\prime}+(1+\omega)(\theta-3\Phi^\prime)=0 \label{dif1},\\
&\theta^\prime+\mathcal{H}(1-3\omega)\theta-\frac{\omega}{(1+\omega)}
k^2 \delta +k^2 \sigma-k^2\Psi=0, \label{dif2}
\end{align}
where $\sigma=\tilde{\sigma} /(1+\lambda \omega^{-1/4})$ is the
same as that of standard cosmology.

\subsection{Impact on the sound horizon and implications for the Hubble tension}
The key mechanism by which EGER addresses the Hubble tension is through its impact on the sound horizon at recombination. The enhanced radiation couplings modify the Hubble parameter at early times, which in turn alters the sound horizon for acoustic waves in the photon–baryon fluid
\begin{eqnarray}
    r_s=\int_{z_{*}}^{\infty}\frac{c_{s\gamma,b}(z)}{H(z)}\, dz,
\end{eqnarray}
where $c_{s\gamma,b}(z)$ is the sound speed and $H(z)$ includes the modified radiation contributions
\begin{align}
H(z)=
100
\sqrt{w_{m0}(1+z)^3+
\sum_{\text{r}}
\left(1+3^{1/4}\lambda_{\text{r}}
\right)w_{\text{r0}}(1+z)^4} 
\quad
\text{km}\, \text{s}^{-1}\, \text{Mpc}^{-1} \label{HP}
\end{align}
with $w_{\text{i}} = \Omega_{\text{i}} h^2$ for each species. The modified radiation sector effectively increases the expansion rate before recombination, thus reducing $r_s$. Since the observed angular scale of the acoustic peaks $\theta_s = r_s / D_A$ is tightly constrained by the CMB, a smaller $r_s$ implies a smaller angular diameter distance $D_A$. Given that $D_A \propto H_0^{-1}$, this naturally leads to a higher inferred value of the Hubble constant, thus helping to reconcile early- and late-universe measurements of $H_0$. However, a quantitative analysis is necessary to examine these points, and this constitutes the focus of the sections that follow.

\section{Analysis of the scale-free model}
\label{sec:analysis}
\subsection{Methodology}
We perform an MCMC analysis using the Einstein-Boltzmann solver \texttt{CLASS} \cite{class} and the MCMC sampler \texttt{MontePython} \cite{montepython1, montepython2}. We also use the simulated-annealing minimiser \texttt{Procoli} \cite{procoli} to find the best-fit or \textit{maximum a posteriori} (MAP) points. The models we will compare are as follows:
\begin{enumerate}
    \item $\Lambda$CDM
    \item $\Lambda$CDM + $N_\text{eff}$
    \item $\Lambda$CDM + $\lambda_\gamma + \lambda_\nu$
\end{enumerate}
where $N_\text{eff}$ is the effective number of relativistic species other than photons and $\{\lambda_\gamma$, $\lambda_\nu\}$ are the scale-free EGER\footnote{henceforth referred to as just EGER.} parameters for photons and neutrinos, respectively. The $N_\text{eff}$ model is a good approximation to models which introduce new relativistic species in the early universe; it thus serves as a good baseline by which to evaluate the performance of EGER, in addition to $\Lambda$CDM. 



For each model, we impose broad uniform priors on the six baseline $\Lambda$CDM parameters and any new parameters specific to the model. We assume three degenerate massive neutrinos with a total mass of $\Sigma m_\nu = 0.06 \text{ eV}$, except when allowing $N_\text{eff}$ to vary, where we make no assumptions about the underlying neutrino distribution. We also use \texttt{hmcode} \cite{hmcode} for nonlinear corrections to the matter power spectrum, \texttt{hyrec} \cite{hyrec} for recombination, and \texttt{PRIMAT} \cite{primat} for BBN calculations. Finally, we use \texttt{GetDist} \cite{getdist} for plotting.

The datasets we use to constrain these models are as follows:
\begin{enumerate}
    \item Low-$\ell$ TT \& EE CMB data from \textit{Planck} PR3 via \texttt{commander} and \texttt{sroll2} \cite{sroll2},
   \item High-$\ell$ TT, TE, \& EE CMB data from \textit{Planck} PR3 via \texttt{plik} \cite{planck}; ACT DR6 \cite{ACT1} and SPT-3G D1 \cite{spt3g} via \texttt{candl} \cite{candl},
    \item CMB lensing data from \textit{Planck} NPIPE, ACT DR6, and SPT-3G D1 \cite{ACT2, ACT3}, 
    \item BAO data from DESI DR2 \cite{desi2, desi1}, 
    \item Pantheon+ SNeIa compilation \cite{pantheonplus}, with and without the SH0ES calibration on the absolute supernovae magnitude $M_b$ \cite{shoes1}.
\end{enumerate}
When combining \textit{Planck} and ACT data, we truncate the \textit{Planck} TT spectrum to $\ell < 1000$ and the TT and TE spectra to $\ell < 600$, as recommended by the ACT collaboration \cite{ACTmain}. This is to limit covariance due to the sky overlap between the two experiments. In particular, we test two dataset combinations: $\mathcal{D}_1$ with \textit{Planck}, ACT, and SPT-3G CMB data, and $\mathcal{D}_2$ with only \textit{Planck} and SPT-3G CMB data (without ACT). BAO and SNeIa data for both combinations are the same. We note that there has been discussion in the literature on the suitability of combining BAO data from DESI DR2 with primary CMB data, as the constraints in the $\Omega_\text{m}$-$hr_\text{d}$ plane from DESI are discrepant with various CMB datasets at around $3\sigma$ \cite{spt3g}. However, we consider this to be beyond the scope of this work and combine these datasets for ease of comparison with results in the literature.

To evaluate the performance of each model in addressing the Hubble tension, we employ three standard tension metrics \cite{olympics1, olympics2}:
\begin{enumerate}
    \item the Gaussian tension $\Delta_\text{{GT}}$ between the model posterior and the SH0ES value for $H_0$, where $H_0 = 73.17 \pm 0.86 \text{ km s}^{-1} \text{ Mpc}^{-1}$ \cite{shoes2}. It is defined as
    \begin{equation}
        \Delta_\text{GT} = \frac{x_\text{model} - x_\text{SH0ES}}{\sqrt{\sigma^2_\text{model} + \sigma^2_\text{SH0ES}}}.
    \end{equation}
    Instead of using $H_0$, one could also use $M_b$ to determine the tension, but we choose to report the tension in $H_0$ for clarity. Moreover, the tension in $M_b$ is roughly equal to the tension obtained with the following metric, and it would be repetitive if we were to report the same values twice.
    \item $Q_\text{DMAP}$, the difference between the \textit{maximum a posteriori} (MAP) points with and without the SH0ES $M_b$ calibration, defined as 
    \begin{equation}
   Q_\text{DMAP} = \sqrt{\chi^2_{\text{min,  w/ SH0ES}} - \chi^2_{\text{min, w/o SH0ES}}} .
    \end{equation}
    In practice, for $\chi^2_\text{min, w/ SH0ES}$ with SH0ES, we run MCMC chains with the \texttt{PantheonPlusSH0ES} likelihood in \texttt{MontePython}. The minimum $\chi^2$ value is then obtained by running the MAP against the baseline dataset with the uncalibrated \texttt{PantheonPlus} SNeIa likelihood plus a SH0ES $M_b$ prior. The value for $Q_\text{DMAP}$ obtained can be shown to be mathematically equivalent to $\Delta_\text{GT}$ evaluated using the SH0ES value for $M_b$ (not $H_0$) for Gaussian posteriors. \cite{olympics2}.  
    \item $\Delta$AIC, the difference in the Akaike Information Criterion with respect to $\Lambda$CDM, defined as
    \begin{equation}
        \Delta \text{AIC} = \chi^2_\text{min, model} - \chi^2_{\text{min}, \  \Lambda\text{CDM}} + 2\mathcal{N},
    \end{equation}
    where $\mathcal{N}$ is the number of extra parameters on top of $\Lambda$CDM. This metric has the added benefit of taking into account model complexity by penalising a large number of extra parameters.
\end{enumerate}

For the first two metrics, we consider a value of $\le 3\sigma$ to be a significant reduction in  tension; for $\Delta \text{AIC}$, we consider a value of $\le 6.91$ to be significant (a more-than-weak preference on Jeffreys' scale).

\subsection{Results}

\subsubsection{EGER vs. $N_\text{eff}$}

From the values in Tables \ref{tab:table1} and \ref{tab:table2}, it can be seen that the ability of EGER to address the Hubble tension is highly correlated with the value of $N_\text{eff}$ preferred by a given dataset. For the dataset $\mathcal{D}_1$ which includes ACT data, the inferred value of $N_\text{eff}$ is slightly lower than the Standard Model (SM) value of $N_\text{eff} = 3.044$, due to the ACT data's preference for increased power in the CMB damping tail \cite{ACTextended}. Since  a higher $N_\text{eff}$ leads to increased Silk damping via a positive degeneracy with $H_0$ \cite{neudamping}, this limits the ability of the $N_\text{eff}$ model to alleviate the Hubble tension. Similarly, EGER also performs poorly, at best reducing the tension to $4.8\sigma$. However, when the inferred value of $N_\text{eff}$ is higher than the SM value, as is the case with the $\mathcal{D}_2$ dataset which does not include ACT data, EGER performs significantly better, being able to reduce the tension to at least $3.9\sigma$ and at best $3.5\sigma$, depending on the metric used. 

\begin{table*}[!h]
\caption{\label{tab:table1} Results for various models under dataset $\mathcal{D}_1$. The values for $H_0$ are quoted as (bestfit) mean $\pm \ 1\sigma$.}
\renewcommand{\arraystretch}{1.4}
\begin{ruledtabular}
\begin{tabular}{cccccccc}
$\mathcal{D}_1$& \multicolumn{2}{c}{\textbf{without SH0ES}} & \multicolumn{1}{c}{\textbf{with SH0ES}} & \multicolumn{4}{c}{\textbf{Tension Metrics}}\\ 
\hline
Model & $\chi^2_\text{min}$ & $H_0$ & $\chi^2_\text{min}$ & $\Delta_\text{GT}$ & $Q_\text{DMAP}$ & $\Delta$AIC$_\text{w/ SH0ES}$ & $\Delta$AIC$_\text{w/o SH0ES}$ \\
\hline
$\Lambda\text{CDM}$ & 3050.84& $(68.22) \ 68.21 \pm 0.26$ & 3084.98 & 5.5$\sigma$ & $5.8\sigma$ & 0.00 & 0.00 \\ 
$N_\text{eff}$& 3046.08& $ (67.85) \ 67.83 \pm 0.68$ & 3072.60 & 4.9$\sigma$ & 5.1$\sigma$ & $-10.38$& $-2.76$\\ 
EGER& 3049.38& $(67.92) \ 67.82 \pm 0.71$& 3075.23& 4.8$\sigma$& 5.1$\sigma$& $-5.75$& 2.54\\ 
\end{tabular}
\end{ruledtabular}
\end{table*}

\begin{table*}[!h]
\caption{\label{tab:table2} Results for various models under dataset $\mathcal{D}_2$.}
\renewcommand{\arraystretch}{1.5}
\begin{ruledtabular}
\begin{tabular}{cccccccc}
$\mathcal{D}_2$& \multicolumn{2}{c}{\textbf{without SH0ES}} & \multicolumn{1}{c}{\textbf{with SH0ES}} & \multicolumn{4}{c}{\textbf{Tension Metrics}}\\ 
\hline
Model & $\chi^2_\text{min}$ & $H_0$ & $\chi^2_\text{min}$ & $\Delta_\text{GT}$ & $Q_\text{DMAP}$ & $\Delta$AIC$_\text{w/ SH0ES}$ & $\Delta$AIC$_\text{w/o SH0ES}$ \\
\hline
$\Lambda\text{CDM}$ & 4385.66& $(68.09) \ 68.08  \pm 0.26$& 4420.82& $5.7\sigma$& $5.9\sigma$& 0.00 & 0.00 \\ 
$N_\text{eff}$& 4380.68& $ (68.69) \ 68.85 \pm 0.80$& 4397.52& 3.7$\sigma$& 4.1$\sigma$& $-21.30$& $-2.98$\\
 $\Delta N_\text{eff}$& 4384.54& $(68.85) \ 69.12 ^{+0.58}_{-0.76}$& 4400.37& 3.7$\sigma$& 4.0$\sigma$& $-18.45$&0.88\\ 
EGER& 4380.94& $(68.93) \ 69.06 \pm 0.82$& 4396.21& 3.5$\sigma$& 3.9$\sigma$& $-20.61$& $-0.72$\\ 
\end{tabular}
\end{ruledtabular}
\end{table*}

\begin{figure}[!h]
\includegraphics[width=0.8\textwidth]{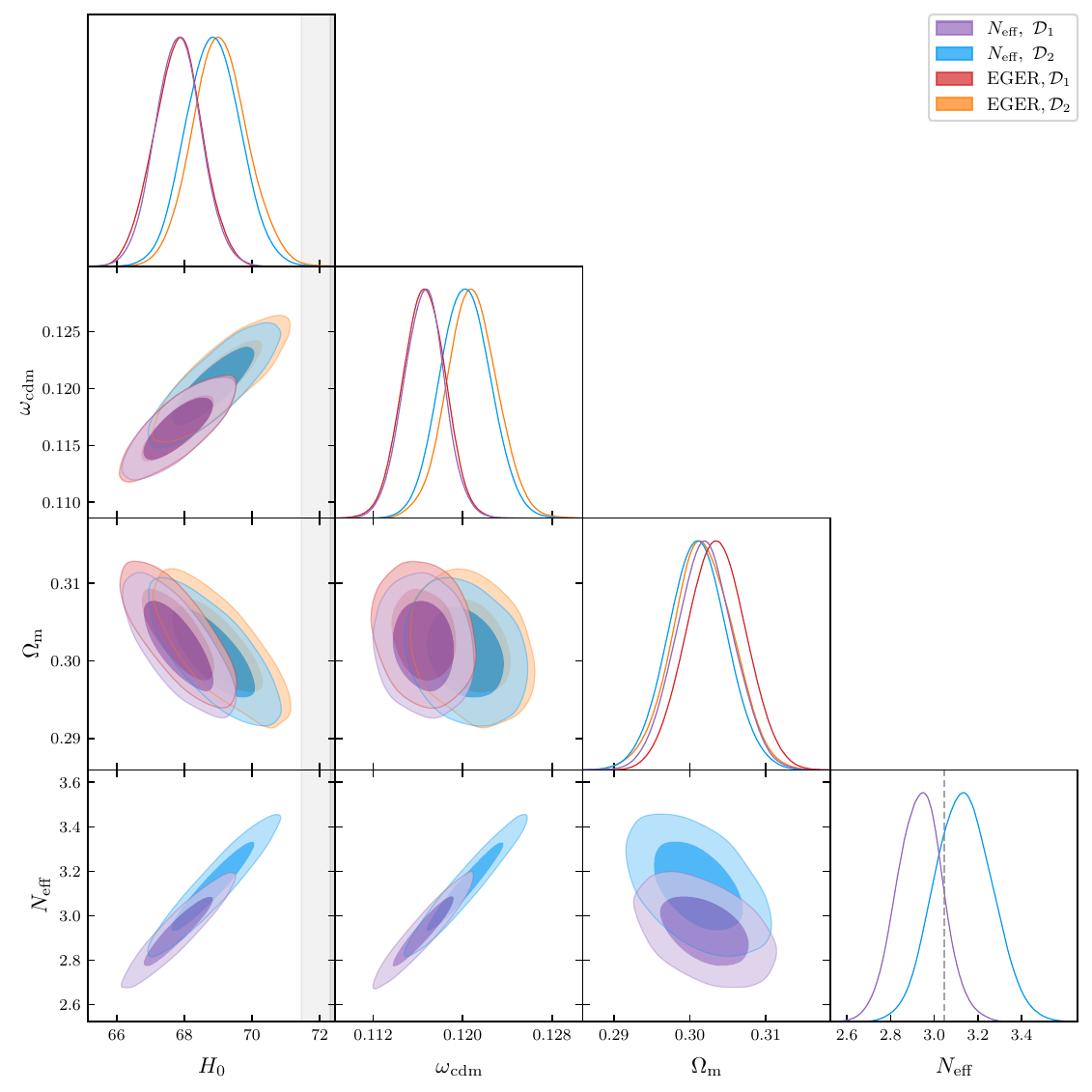}
\caption{\label{fig:fig1} 68\% and 95\% contours for the $N_\text{eff}$ and EGER models. The grey bands denote the 68\% and 95\% contours for the SH0ES measurement of $H_0 = 73.17 \pm 0.86 \text{ km s}^{-1} \text{ Mpc}^{-1}$ \cite{shoes2}, while the grey dashed line represents the SM value of $N_\text{eff} = 3.044$.}
\label{tab: Fig. 1}
\end{figure}

\begin{figure}[!h]
\includegraphics[width=0.9\textwidth]{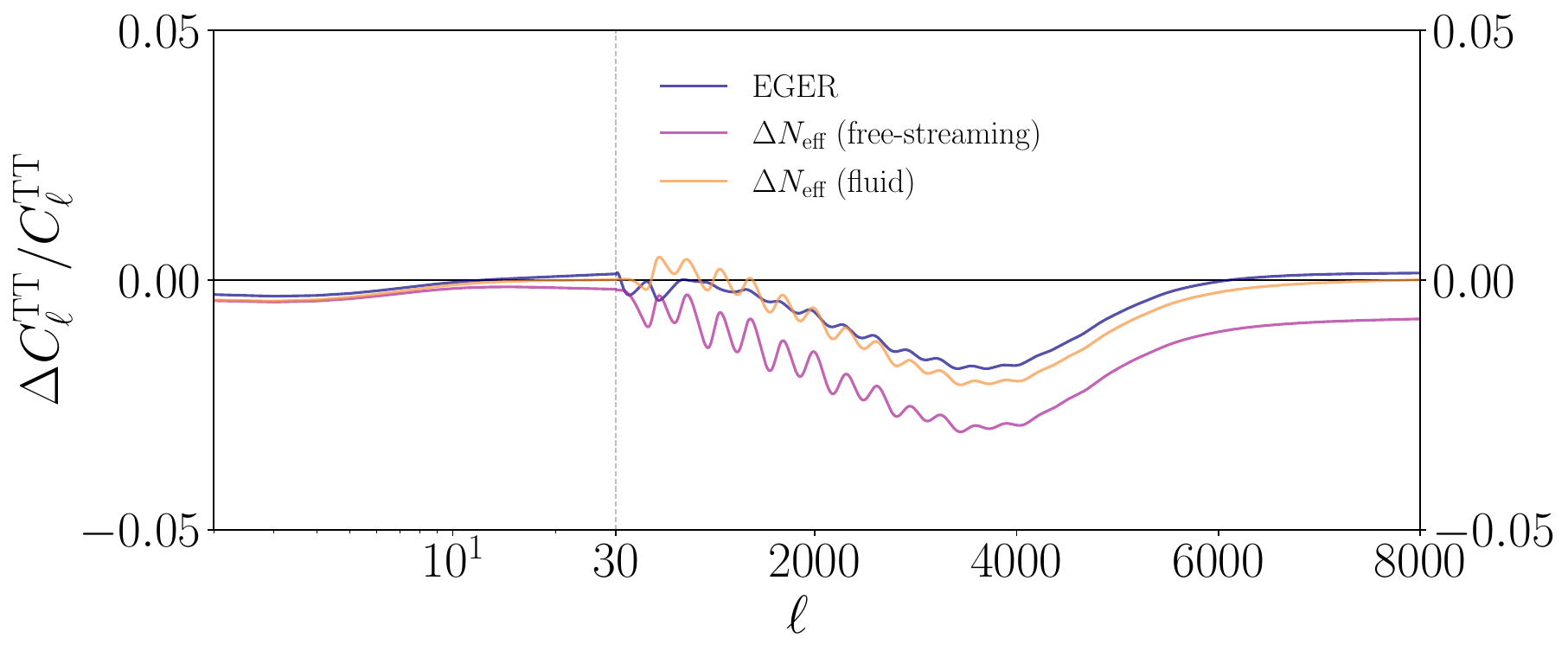}
\caption{\label{fig:fig2} CMB TT power spectrum residuals with respect to the bestfit $\Lambda$CDM spectra. The blue residual is for the bestfit EGER model to dataset $\mathcal{D}_2$ while the other two residuals are for free-streaming and fluid-like dark radiation models with $N_\text{eff} =N_\text{eq}^{\text{bestfit}}$. All three models assume massive neutrinos and all other cosmological parameters are kept constant. }
\label{tab: Fig. 2}
\end{figure}

One can calculate an equivalent $N_\text{eff}$, denoted $N_\text{eq}$, which corresponds to certain values of the EGER parameters $\lambda_\gamma$ and $\lambda_\nu$ as
\begin{equation}
\label{Neq}
    N_{\text{eq}} = 3^{1/4}\left(\frac{8}{7}\right)\left(\frac{4}{11}\right)^{-4/3}\lambda_\gamma + (1 + 3^{1/4}\lambda_\nu)N_\text{eff}^{\text{SM}}.
\end{equation}
For the $N_\text{eff}$ model under dataset $\mathcal{D}_2$, the bestfit value for $N_\text{eff}=3.102$. The bestfit values for $\lambda_\gamma$ and $\lambda_\nu$ for the EGER model under the same dataset are $\lambda_\gamma = 0.040$ and $\lambda_\nu = -0.027$, respectively. Hence, the equivalent $N_{\text{eq}}$ using Eq. \ref{Neq} is $N_{\text{eq}} = 3.168$, somewhat higher than that in the $N_\text{eff}$ model. This contributes to EGER having a slight edge over the $N_\text{eff}$ model when it comes to addressing the Hubble tension, which can be seen via the tension metric values in Tables \ref{tab:table1} and \ref{tab:table2}, but also via the 2D posteriors in Fig. \ref{fig:fig1}, where the two models are otherwise almost indistinguishable from one another but for the slight `edge'. 


To account for the possibility that this advantage is due to the assumption of massive neutrinos, we also tested an additional variation of the $N_\text{eff}$ model, denoted $\Delta N_\text{eff}$, which assumes three massive degenerate neutrinos with a total mass of $\Sigma m_\nu = 0.06 \text{ eV}$. We only test this variant for dataset $\mathcal{D}_2$ as the Bayesian posterior for $N_\text{eff}$ under dataset $\mathcal{D}_1$ peaks at values less than the SM value of $N_\text{eff} = 3.044$, resulting in unusually tight constraints on any extra free-streaming radiation species. From Table \ref{tab:table2}, it can be seen that the effects of this assumption on the ability of the model to address the Hubble tension are negligible. The inferred $N_\text{eff}$ is higher than that without said assumption, with the bestfit value being $N_\text{eff} = 3.193$, even higher than the EGER bestfit value of $N_\text{eq} = 3.168$. From the $\Delta$AIC values, the inclusion of massive neutrinos seems to worsen the fit to the data for the $\Delta N_\text{eff}$ model, while EGER performs similarly to the case where neutrinos are massless\footnote{At least, for dataset $\mathcal{D}_2$. For dataset $\mathcal{D}_1$ where a lower $N_\text{eff}$ than the SM value is preferred, EGER performs similarly to the baseline $\Lambda$CDM model. However, it is still able to alleviate the Hubble tension somewhat compared to $\Lambda$CDM.}. This is expected as models resembling $\Lambda$CDM have been shown to prefer close-to-zero or even negative values for the total neutrino mass when it is allowed to vary in MCMC, which is still a highly discussed issue in the literature \cite{herold, negneu}. 

The main difference between EGER and other $N_\text{eff}$ is at the level of perturbations. At the background level, an increase in energy density whether due to extra radiation species or EGER increases the expansion rate during the radiation-dominated era, leading to a smaller physical comoving sound horizon and a higher inferred $H_0$. However, extra free-streaming radiation species contribute anisotropic stress, leading to a decrease in amplitude of the CMB spectra at scales within the sound horizon, potentially in conflict with measurements \cite{seljak, lesgourguesneutrino}. In the literature, one way around this is with a strongly self-interacting dark radiation (SIDR) species which boosts the expansion rate at the background level but does not contribute to anisotropic stress at the level of perturbations \cite{neuwallisch, sidr}. Similarly, EGER also does not contribute anisotropic stress on its own but merely rescales the contribution of existing radiation species. In Figure \ref{fig:fig2}, we plot CMB temperature power spectrum residuals with respect to $\Lambda$CDM for the bestfit EGER and equivalent $\Delta N_\text{eff}$ and SIDR models (equivalent in the sense that $N_\text{eff} = N_\text{eq}^{\text{bestfit}}$), all assuming massive neutrinos. The residuals for EGER more closely resembles that of SIDR, due to their lack of significant additional anisotropic stress. 

Last but not least, care must be taken when considering the implications of new radiation-gravity couplings for BBN. An enhanced expansion rate during BBN modifies the light-element abundances, in particular the helium fraction $Y_\text{p}$ which plays an important role in shaping the CMB damping tail. $N_\text{eff}$ and SIDR models can sidestep this by simply assuming that the extra radiation is produced after BBN — for example, due to a dark radiation-matter decoupling \cite{DRMD}.

In summary, the scale-free EGER model performs better than the baseline $N_\text{eff}$ extension to $\Lambda$CDM at addressing the Hubble tension, especially when massive neutrinos are assumed. However, its ability to alleviate the tension is highly dependent on whether a higher $N_\text{eff}$ is preferred by a given dataset, as is the case with many models that introduce additional radiation species. This may not apply to EGER models which are not scale-free, but we leave the general case for a future work. Upcoming data from experiments such as SPT-3G, DESI, and \textit{Euclid} will be important in constraining such solutions to the Hubble tension, including EGER and its variants. 

\FloatBarrier
\section{Concluding remarks and prospects}
\label{sec:conclusion}
In this work, we revisited the class of gravity–matter couplings constructed from the determinant of the matter energy–momentum tensor, originally introduced in the astrophysical context of compact stars by one of us and a collaborator \cite{azridet}, and examined their implications for cosmology. The central feature of this construction is the dependence on the determinant of the stress-energy tensor, which is particularly sensitive to pressure. As a result, the modification becomes relevant in relativistic regimes, while leaving nonrelativistic epochs largely unaffected.

We showed that, in the early universe, the stress-energy-determinant coupling selectively influences the radiation-dominated era, providing a correction to the early expansion rate without invoking additional fields beyond the Standard Model particle content. This distinguishes the framework from other stress-energy–based extensions, such as couplings to its trace or its quadratic terms which generically affect both relativistic and nonrelativistic components  \cite{traceT,Tsquare,Tsquare2,Tsquare3}. The determinant structure therefore offers a natural mechanism to alter early-universe dynamics while preserving the standard cosmological evolution at later times.

We presented a detailed derivation of both the background evolution and the linear perturbation dynamics within the framework we refer to as \textit{enhanced gravitational effects of radiation}. As a specific realization, we confronted the scale-free model with current cosmological data from \textit{Planck}, ACT, SPT-3G, DESI DR2, Pantheon+, and lensing measurements via a full Markov Chain Monte Carlo analysis. The resulting constraints demonstrate that the enhanced radiation-gravity couplings provide a statistically viable fit to all datasets and yield a modest reduction of the Hubble tension, comparable to scenarios featuring additional fluid-like radiation. The preferred region of parameter space remains compatible with BBN limits and exhibits a clear degeneracy structure that interpolates between $N_\text{eff}$ and SIDR models.

Future work may explore extensions beyond the scale-free framework. It will also be worthwhile to develop more detailed predictions for upcoming CMB and BAO surveys, which have the sensitivity to distinguish this gravitational mechanism from both free-streaming and fluid-like dark radiation.

\section*{Acknowledgments} H.A. is grateful for discussion to Deanna C. Hooper, Fabian Schmidt, Glenn Starkman and Elham Nazari. A.T. is grateful to Laura Herold for discussion and help with the DESI DR2 likelihoods.

\bibliography{hubble}
\bibliographystyle{unsrt}

\end{document}